\documentclass[aps,pra,letterpaper,twocolumn,showpacs,,amsmath,amssymb,superscriptaddress]{revtex4-1}
\usepackage{graphicx}
\usepackage{setspace}
\usepackage{dcolumn}
\usepackage{mathrsfs}
\usepackage{bm}
\usepackage{longtable}

\begin{document}

\preprint{}

\title{Analyzing a single-laser repumping scheme for efficient loading of a strontium magneto-optical trap}

\author{Fachao Hu}

\thanks{These authors contributed equally to this work.}

\affiliation{Hefei National Laboratory for Physical Sciences at the Microscale and Shanghai Branch, University of Science and Technology of China, Shanghai 201315, China}

\affiliation{CAS Center for Excellence and Synergetic Innovation Center in Quantum Information and Quantum Physics, University of Science and Technology of China, Shanghai 201315, China}

\author{Ingo Nosske}

\thanks{These authors contributed equally to this work.}

\affiliation{Hefei National Laboratory for Physical Sciences at the Microscale and Shanghai Branch, University of Science and Technology of China, Shanghai 201315, China}

\affiliation{CAS Center for Excellence and Synergetic Innovation Center in Quantum Information and Quantum Physics, University of Science and Technology of China, Shanghai 201315, China}

\author{Luc Couturier}

\affiliation{Hefei National Laboratory for Physical Sciences at the Microscale and Shanghai Branch, University of Science and Technology of China, Shanghai 201315, China}

\affiliation{CAS Center for Excellence and Synergetic Innovation Center in Quantum Information and Quantum Physics, University of Science and Technology of China, Shanghai 201315, China}

\author{Canzhu Tan}

\affiliation{Hefei National Laboratory for Physical Sciences at the Microscale and Shanghai Branch, University of Science and Technology of China, Shanghai 201315, China}

\affiliation{CAS Center for Excellence and Synergetic Innovation Center in Quantum Information and Quantum Physics, University of Science and Technology of China, Shanghai 201315, China}

\author{Chang Qiao}

\affiliation{Hefei National Laboratory for Physical Sciences at the Microscale and Shanghai Branch, University of Science and Technology of China, Shanghai 201315, China}

\affiliation{CAS Center for Excellence and Synergetic Innovation Center in Quantum Information and Quantum Physics, University of Science and Technology of China, Shanghai 201315, China}

\author{Peng Chen}


\affiliation{Hefei National Laboratory for Physical Sciences at the Microscale and Shanghai Branch, University of Science and Technology of China, Shanghai 201315, China}

\affiliation{CAS Center for Excellence and Synergetic Innovation Center in Quantum Information and Quantum Physics, University of Science and Technology of China, Shanghai 201315, China}

\author{Y. H. Jiang}

\email{jiangyh@sari.ac.cn}

\affiliation{Shanghai Advanced Research Institute, Chinese Academy of Sciences, Shanghai 201210, China}

\affiliation{CAS Center for Excellence and Synergetic Innovation Center in Quantum Information and Quantum Physics, University of Science and Technology of China, Shanghai 201315, China}

\author{Bing Zhu}

\email{bzhu@physi.uni-heidelberg.de}

\affiliation{Physikalisches Institut, Universit\"at Heidelberg, Im Neuenheimer Feld 226, 69120 Heidelberg, Germany}

\affiliation{Hefei National Laboratory for Physical Sciences at the Microscale and Shanghai Branch, University of Science and Technology of China, Shanghai 201315, China}

\affiliation{CAS Center for Excellence and Synergetic Innovation Center in Quantum Information and Quantum Physics, University of Science and Technology of China, Shanghai 201315, China}

\author{Matthias Weidem\"uller}

\email{weidemueller@uni-heidelberg.de}

\affiliation{Hefei National Laboratory for Physical Sciences at the Microscale and Shanghai Branch, University of Science and Technology of China, Shanghai 201315, China}

\affiliation{CAS Center for Excellence and Synergetic Innovation Center in Quantum Information and Quantum Physics, University of Science and Technology of China, Shanghai 201315, China}

\affiliation{Physikalisches Institut, Universit\"at Heidelberg, Im Neuenheimer Feld 226, 69120 Heidelberg, Germany}

\date{\today}

\begin{abstract}
We demonstrate enhanced loading of strontium atoms into a magneto-optical trap using a repumping scheme from the metastable state via the doubly-excited state $5\mathrm{s}5\mathrm{p}\,^3\mathrm{P}_2 \rightarrow 5\mathrm{p}^2\,^3\mathrm{P}_2$ at $481~\mathrm{nm}$. The number of trapped atoms is increased by an order of magnitude. The frequency and intensity dependence of the atom number enhancement, with respect to the non-repumping case, is well reproduced by a simple rate equation model, which also describes single-laser repumping schemes reported previously. The repumping scheme is limited by a weak additional loss channel into the long-lived $5\mathrm{s}5\mathrm{p}\,^3\mathrm{P}_0$ state. For low repumping intensities, the signature of a halo formed by magnetically trapped atoms in the metastable state is found.
\end{abstract}

\maketitle

\section{Introduction}
\begin{figure}[t!]
  \centering
  \includegraphics[width=0.95\linewidth]{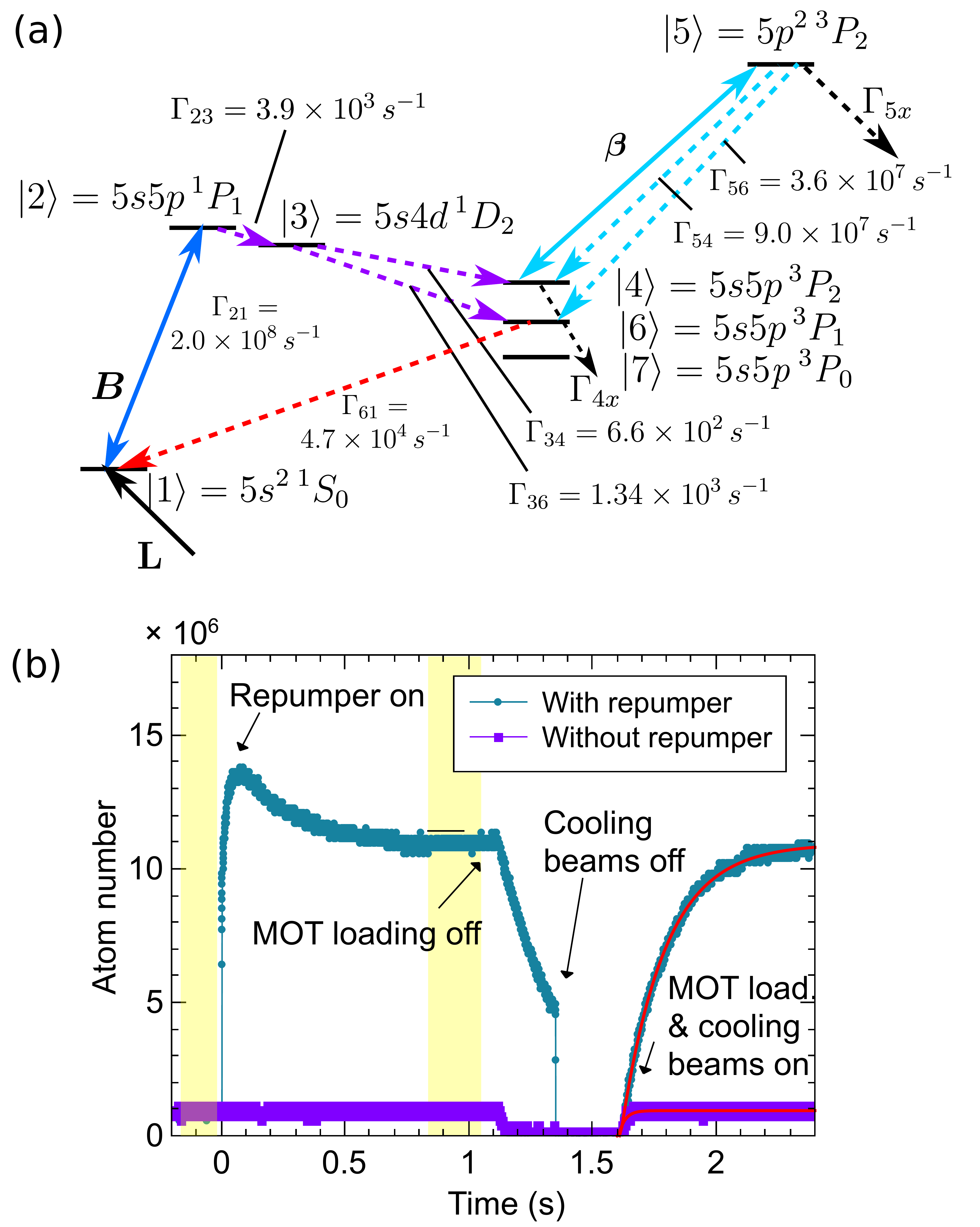}
  \caption{(a) Relevant levels, transition and loss channels of strontium for laser cooling and repumping. (b) $^{88}$Sr MOT fluorescence signal with and without repumping. With repumping, a burst signal appears at $t = 0$ when the repumper is turned on. The steady-state MOT fluorescence signals in the yellow regions are for computing atom number enhancement. The loading curves after $1.6~ \mathrm{s}$ fit very well to simple exponentials (red lines).}
\label{fig:fig1}
\end{figure}

The interest in ultracold strontium atoms stems from diverse research areas such as high-precision metrology \citep{ludlow2015, campbell2017}, the quest for a continuous atom laser \citep{bennetts2017}, molecular physics \citep{mcguyer2015} or quantum simulation with two-electron Rydberg atoms \citep{dunning2016}. For these applications it is often desirable to provide large numbers of cold atoms. One obstacle in achieving large atom numbers, however, is considerable loss during the first trapping and cooling stage of strontium. This is elucidated in Fig. \ref{fig:fig1}(a), in which the relevant electronic levels and transitions of atomic strontium are shown. Following standard laser cooling techniques, the atoms are captured in a three-dimensional magneto-optical trap (3D-MOT) using the broad $5\mathrm{s}^2 \, ^1\mathrm{S}_0 \rightarrow 5\mathrm{s}5\mathrm{p} \, ^1\mathrm{P}_1$ transition at the wavelength of 461 nm \citep{kurosu1990, stellmer2014a}. While this transition connects a nearly closed two-level system, there is the weak decay channel $5\mathrm{s}5\mathrm{p} \, ^1\mathrm{P}_1 \rightarrow 5\mathrm{s}4\mathrm{d} \, ^1\mathrm{D}_2 \rightarrow 5\mathrm{s}5\mathrm{p} \, ^{3}\mathrm{P}_{2}$. The metastable $5\mathrm{s}5\mathrm{p} \, ^3\mathrm{P}_2$ state has a lifetime on the order of 100 s in a room-temperature ultra-high vacuum environment, limited by blackbody coupling to nearby subsequently decaying $5\mathrm{s}4\mathrm{d} \, ^3\mathrm{D}_J$ states (not shown in Fig. \ref{fig:fig1} (a)) \citep{yasuda2004}. This decay channel limits the storage time of a MOT to roughly 10 ms and, thus, the number of trapped atoms \citep{dinneen1999,xu2003}.

To overcome this limitation and to enhance the atom number in a strontium MOT, several schemes have been realized to repump atoms in the $5\mathrm{s}5\mathrm{p} \, ^3\mathrm{P}_2$ state. In early experiments, the repumping scheme for strontium addresses the two $5\mathrm{s}5\mathrm{p} \, ^3\mathrm{P}_{0,2} \rightarrow 5\mathrm{s}6\mathrm{s} \, ^3\mathrm{S}_1$ transitions at 679 nm and 707 nm, respectively \citep{dinneen1999}. The laser at 679 nm removes atoms from the $5\mathrm{s}5\mathrm{p} \, ^3\mathrm{P}_0$ state and closes the cycle \cite{xu2003}. Alternatively, single-laser repumping schemes addressing transitions to higher lying states $5\mathrm{s}5\mathrm{p} \, ^3\mathrm{P}_2 \rightarrow 5\mathrm{s}n\mathrm{d} \, ^3\mathrm{D}_2$ were demonstrated, specifically using $n=4$ at 3012 nm \citep{mickelson2009}, $n=5$ at 497 nm \citep{poli2005, poli2006, stellmer2014b, moriya2018,Schkolnik2018} and $n=6$ at 403 nm \citep{stellmer2014b, moriya2018}. Recently, repumping directly from $5\mathrm{s}4\mathrm{d} \, ^1\mathrm{D}_{2}$ using a 448 nm diode laser has been proposed \citep{Mills2017} and realized \cite{schrecPriv}. An increase of the steady-state atom number by a factor of up to 20 \citep{mickelson2009} and 30 \citep{poli2005, stellmer2014b}, at 3012 nm and 497 nm respectively, has been realized by such schemes employing only a single repumping transition.

We report on the realization of a simple repumping scheme of strontium atoms in the metastable $5\mathrm{s}5\mathrm{p} \, ^3\mathrm{P}_2$ state via the doubly-excited $5\mathrm{p}^2 \, ^3\mathrm{P}_2$ state at the wavelength of $481.323~\mathrm{nm}$ delivered by a diode laser. Repumping via this state has been realized earlier \cite{Camargo2017}, but a thorough characterization of this scheme is lacking. We measure the dependence of atom number enhancement on repumping frequency and intensity, and identify the major limitations of this repumping scheme by comparing the measurements with a simple rate equation model. In addition, we apply this model to alternative repumping schemes demonstrated earlier.

\section{Characterization of the repumping scheme}

\subsection{Measurements}

Atoms are loaded from a two-dimensional MOT (2D-MOT) \citep{nosske2017} and subsequently trapped in a 3D-MOT, created by three retro-reflected cooling beams at 461 nm with a total peak intensity of $40~\mathrm{mW/cm}^2$, a detuning of $-30~\mathrm{MHz}$ and a $1/e^2$ radius of $6~\mathrm{mm}$. The axial magnetic field gradient is $33~\mathrm{G/cm}$. The MOT cloud diameter is about $1~\mathrm{mm}$ under these conditions. The repumping light is generated from an external-cavity diode laser (ECDL, Toptica DL pro), with a maximum output power of more than 10 mW which is fiber coupled into the main chamber. The repumping beam has a $1/e^2$ radius of $2.5~\mathrm{mm}$. Both the cooling and the repumping lasers are frequency-stabilized using a commercial wavelength meter \citep{couturier2018}. For the measurement presented here, the frequency of the repumping light is tuned by scanning the piezo of the ECDL, and the power is changed by tuning the fiber coupling efficiency. The MOT atom number is $10^6$ without the repumping light. 

Shown in Fig. \ref{fig:fig1} (a) are the relevant levels and rates of strontium during laser cooling and repumping. Values for decay rates $\Gamma_{ij}$ are taken from Ref. \citep{sansonetti2010,xu2003}. $B$ and $\beta$ denote the pumping rates on the cooling and the repumping transitions, $L$ is the MOT loading rate and $\Gamma_{ix}$ denotes decays out of the system from a state $|i\rangle$. Atoms in $5\mathrm{s}5\mathrm{p} \, ^1\mathrm{P}_1$ have a probability of $1 {:} 50{,}000$ \citep{xu2003,newRatio} decaying to $5\mathrm{s}4\mathrm{d} \, ^1\mathrm{D}_2$ state, from which they decay further to $5\mathrm{s}5\mathrm{p} \, ^{3}\mathrm{P}_{1,2}$ with a branching ratio of $2{:}1$. Atoms in $5\mathrm{s}5\mathrm{p} \, ^{3}\mathrm{P}_{2}$ are repumped to the doubly-excited state $5\mathrm{p}^2 \, ^3\mathrm{P}_{2}$ which subsequently decays to the $5\mathrm{s}5\mathrm{p} \, ^{3}\mathrm{P}_{1,2}$ states with a branching ratio of $\sim 3 {:} 1$ \citep{garcia1988,sansonetti2010}. The short-lived $5\mathrm{s}5\mathrm{p} \, ^{3}\mathrm{P}_{1}$ state atoms decay to the ground state and do not result in atom loss, which is even beneficial for the repumping. There are no other dipole-allowed single-photon decay channels. Meanwhile, the loss rates $\Gamma_{4x}$ and $\Gamma_{5x}$ from states $|4\rangle$ and $|5\rangle$ during repumping have to be taken into account, due to e.g. collisional loss or a small branching ratio to the long-lived $5\mathrm{s}5\mathrm{p} \, ^{3}\mathrm{P}_{0}$ state, as shown in Fig. \ref{fig:fig1} (a).

To characterize the enhancement, we compare the steady-state MOT atom numbers with and without repumping light at various intensities and detunings. For each measurement of different repumping intensities and frequencies, the MOT is operated for a fixed time of 2 s during which the atoms in the $5\mathrm{s}5\mathrm{p} \, ^{3}\mathrm{P}_{2}$ state are accumulated in a magnetic trap formed by the MOT gradient field \cite{stellmer2014b,xu2003}. The repumping beam is turned on at $t=0$ (see the sequence in Fig. \ref{fig:fig1} (b)), resulting in a sudden fluorescence increase as atoms in the reservoir state are pumped back to the MOT cycle, followed by a fluorescence decay before atom populations reach a steady state. The atom number enhancement $\mathcal{E}$ is defined as the ratio of the steady-state atom numbers in the ground state with and without repumping, see the yellow shaded regions in Fig \ref{fig:fig1} (b).  The atom flux is stopped \cite{nosske2017} at $t=1.15~\mathrm{s}$, and the cooling beams are extinguished at $t=1.35~\mathrm{s}$. At $t=1.6~\mathrm{s}$ the flux and the cooling beams are turned on and MOT loading curves are recorded to show the daily operation. These loading curves are well fitted with simple exponentials, $N = \frac{L}{R} [1-\exp(-Rt)]$ with the loss rate $R$.

\begin{figure}[t!]
  \centering
  \includegraphics[width=0.8\linewidth]{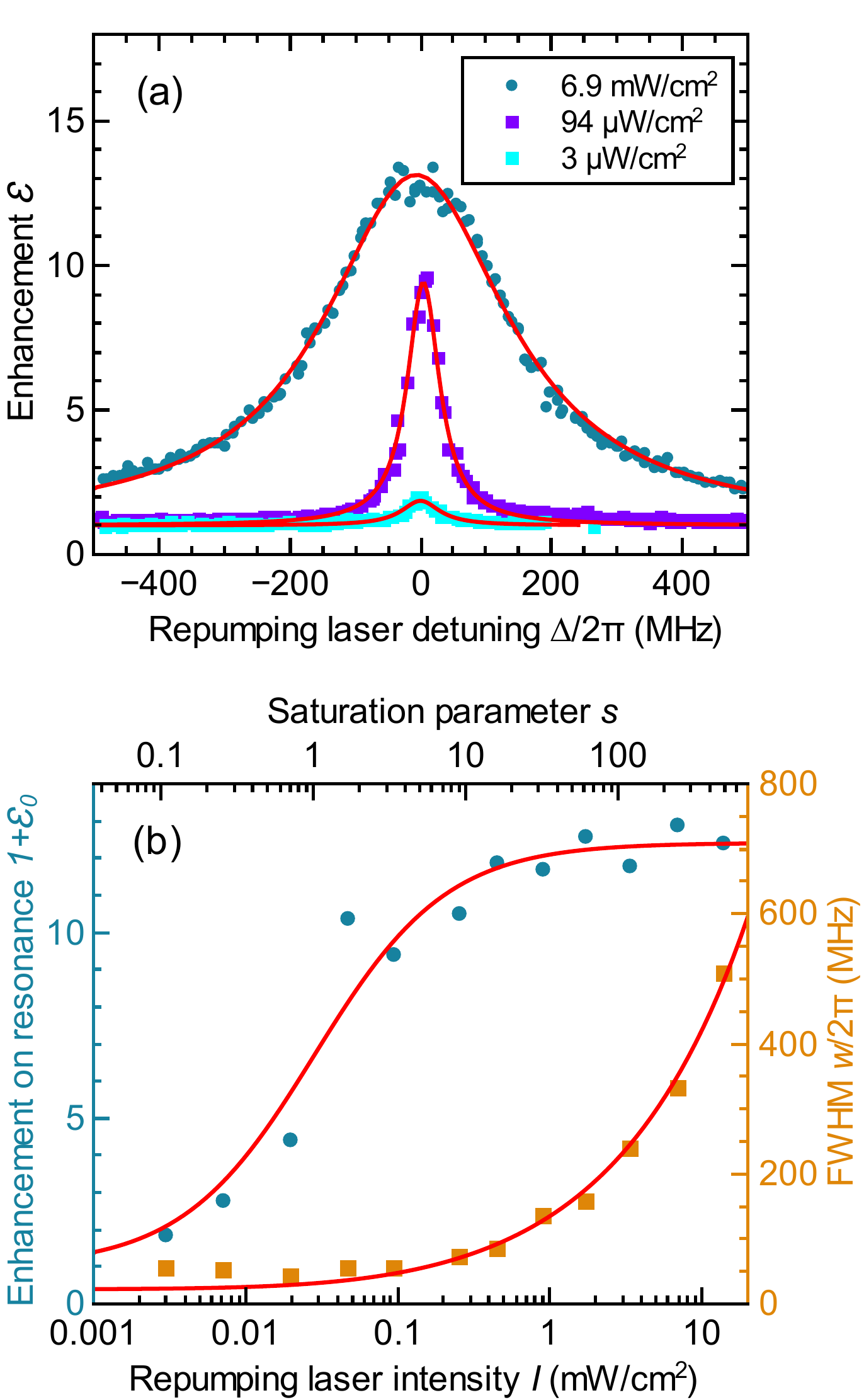}
  \caption{Enhancement of the MOT atom number due to the repumping light. (a) Dependence of the enhancement on the repumping laser frequency for different repumping beam intensities. The red lines are Lorentzians fitted to the data. (b) Enhancement on resonance (blue circles) and the FWHM (yellow squares) of the fitted Lorentzians versus intensity of the repumping laser. The red lines are fits to the solutions of a rate equation model as explained in the text.}
\label{fig:fig2}
\end{figure}

\subsection{Analysis}

The enhancement for different repumping beam detuning $\Delta$ and mean intensity $I$ is shown in Fig. \ref{fig:fig2} (a). Each enhancement spectrum at a fixed intensity is fitted by a Lorentzian of the form $\mathcal{E} = 1 + \mathcal{E}_0 \, \frac{w^2}{4\Delta^2 + w^2}$, with the amplitude $\mathcal{E}_0$ and the full width at half maximum (FWHM) $w$. This function appears as the steady-state solution of a simple rate equation model which takes the six levels of Fig. \ref{fig:fig1} (a) into account (see the Appendix for its derivation). The unknown loss rates $\Gamma_{4x}$ and $\Gamma_{5x}$ are included to account for additional loss channels. From the rate equation model follows that the intensity dependence of $\mathcal{E}_0$ is given by $\mathcal{E}_0 = \frac{\Gamma_{56}}{\Gamma_{5x}} \frac{s}{1 + s}$, with the saturation parameter $s = I/I_s$. $I_s$ being the effective saturation intensity (see Eq. \ref{eq:Isat}). We fix $\Gamma_{56} = 3.6 \times 10^7~\mathrm{s}^{-1}$ according to the literature value \citep{sansonetti2010}. By fitting the measured on-resonance enhancement $1+\mathcal{E}_0$, we deduce the loss rate from the doubly-excited state as $\Gamma_{5x} = 3.2(1) \times 10^6~\mathrm{s}^{-1}$ and the effective saturation intensity as $I_s = 28(6)~\mu\mathrm{W/cm}^2$. With $\Gamma_{5x}$ and $I_s$ known, $\Gamma_{4x} \sim 10^{3}~\mathrm{s}^{-1}$ is computed from Eq. (\ref{eq:Isat}). this value is much larger than the vacuum-limited loss rate (about 0.1 $\mathrm{s}^{-1}$ deduced from the magnetic trapping of the atoms), indicating binary collisional losses involving atoms in excited states through, e.g., Penning ionization. This interpretation by binary losses is supported by the observation that the enhancement becomes larger for smaller atom densities \cite{mickelson2009,stellmer2014b}. By decreasing the 2D MOT atom flux \citep{nosske2017}, the blue MOT atom number without repumping light is reduced to $10^4$ with a much lower density, and the enhancement with repumping light is as high as 20.

The rate equation model also properly reproduces the measured dependence of the linewidth on the repumping light intensity. We fit the data by $w = \Gamma_5 \sqrt{1+s}$, with the total decay rate from the doubly-excited state $\Gamma_5 = \Gamma_{54} + \Gamma_{56}$ as the only free parameter. The fitted value $\Gamma_5 = 2\pi \times 22(3)~\mathrm{MHz}$, is consistent with the literature value, $\Gamma_{5} = \Gamma_{54} +  \Gamma_{56} = 2\pi \times 20(1)~\mathrm{MHz}$ \citep{sansonetti2010}.

For low repumper intensities the measured linewidth of about 50 MHz lies above the prediction of the rate equation model. In this regime, a significant fraction of atoms is pooled in the metastable state $5\mathrm{s}5\mathrm{p} \, ^3\mathrm{P}_2$. While these atoms do not experience the light force of the MOT, they are magnetically trapped by the quadrupole field of the MOT coils, thus forming a halo around the optically trapped atomic cloud. Eventually, these atoms decay back into the ground state and contribute to the fluorescence signal with a Zeeman broadening corresponding to the spatial distribution of the halo (see also Ref. \cite{moriya2018}). From the MOT temperature of a few mK and our magnetic field gradient, the halo radius is estimated to be around 1 cm, which results in a Zeeman shift consistent with the measured linewidth. In fact, the linewidth measurements shown in Fig. 3(c) in Ref. \cite{stellmer2014b} might actually also be caused by this halo effect.

The simple rate equation model employed here can be applied to other similar repumping schemes. As an example, we have analyzed the repumping scheme described in Ref. \citep{moriya2018}, which addresses the $5\mathrm{s}5\mathrm{d} \, ^3\mathrm{D}_2$ state (see Fig. 2 (c) of Ref. \citep{moriya2018} and Fig. 3 (c) of Ref. \cite{stellmer2014b}, respectively). Fitting the steady-state enhancement curve in Ref. \citep{moriya2018} results in a maximum enhancement of 12, in good agreement with the measured value. Concerning the finding in Ref. \cite{stellmer2014b}, the results of a reservoir spectroscopy approach are modeled by a simplified set of rate equations where only $|4\rangle$, $|5\rangle$ and $|6\rangle$ are involved. A time-dependent solution, using the experimental parameters, reproduces the spectroscopy amplitude and Lorentzian linewidth dependence on the repumping intensity very well. The saturation parameter is found to agree within 1\% to the reported value in the paper.

\section{Conclusion}

In conclusion, we have realized a simple, yet efficient repumping scheme enhancing atom numbers in a strontium MOT by more than one order of magnitude as compared to the performance without any repumping laser. As such, the efficiency of this scheme is comparable to previously reported repumping schemes using a single laser source \citep{poli2005, mickelson2009, stellmer2014b}. We load $10^7$ $^{88}$Sr atoms into the  MOT at a loading rate of $10^8~\mathrm{s}^{-1}$. We identify spurious decay of the doubly-excited state, eventually populating the $5\mathrm{s}5\mathrm{p} \, ^3\mathrm{P}_0$ state, as the main source for residual loss. As a consequence, the enhancement may be further increased by adding another laser repumping atoms from the $5\mathrm{s}5\mathrm{p} \, ^3\mathrm{P}_0$ state, e.g., via the $5\mathrm{p}^2 \, ^3\mathrm{P}_1$ state at the wavelength of 474 nm \citep{sansonetti2010}. This would result in an enhancement of the order of 100 \cite{stellmer2014b}, then finally limiting the final atom number in the strontium MOT by collisional loss \cite{dinneen1999}. We find the halo effect at low repumping intensities by comparing the measurement with the rate equation model. Our rate equation model can be generalized to describe two-laser repumping schemes.

\section*{Acknowledgments}

We thank R. Celistrino Teixeira, S. Stellmer and F. Schreck for providing the original data of their papers and for helpful discussions. M.W.'s research activities in China are supported by the 1000-Talent-Program of the Chinese Academy of Sciences. The work was supported by the National Natural Science Foundation of China (Grant Nos. 11574290 and 11604324) and Shanghai Natural Science Foundation (Grant No. 18ZR1443800). Y.H.J. also acknowledges support under Grant Nos. 11420101003 and 91636105. P.C. acknowledges support of Youth Innovation Promotion Association, the Chinese Academy of Sciences.

\section*{Appendix}
\setcounter{equation}{0}
\renewcommand\theequation{A.\arabic{equation}}

We use a set of rate equations to describe the populations of all six relevant states during the repumping process, as shown in Fig. \ref{fig:fig1} (a):
\begin{eqnarray}
& &\dot{N}_1 = L + B (N_2 - N_1) + \Gamma_{21} N_2 + \Gamma_{61} N_6~,\nonumber\\
& &\dot{N}_2 = -B (N_2 - N_1) - (\Gamma_{21}+\Gamma_{23}) N_2~,\nonumber\\
& &\dot{N}_3 = \Gamma_{23} N_2 - (\Gamma_{34}+\Gamma_{36}) N_3~,\nonumber\\
& &\dot{N}_4 = \beta (N_5 - N_4) + \Gamma_{34} N_3 + \Gamma_{54} N_5 - \Gamma_{4x} N_4~,\nonumber\\
& &\dot{N}_5 = -\beta (N_5 - N_4) - (\Gamma_{54}+\Gamma_{56}+\Gamma_{5x}) N_5~,\nonumber\\
& &\dot{N}_6 = \Gamma_{36} N_3 + \Gamma_{56} N_5 - \Gamma_{61} N_6~.
\end{eqnarray}

\noindent As in the sub-system of states $|1\rangle$ and $|2\rangle$, $\Gamma_{21}$ and $B$ (on the order of $\Gamma_{21}$) are much larger than the other rates, we can adiabatically eliminate their time dependences and find $N_1 = \frac{B + \Gamma_{21}}{B} N_2$. The steady-state solution for the atom number in state $|2\rangle$, to which the measured fluorescence signal is proportional, is given by
\begin{eqnarray}
N_2^{ss} = \frac{L}{\Gamma_{23}} \frac{\Gamma_{34}+\Gamma_{36}}{\Gamma_{34}} \, \mathcal{E},
\end{eqnarray}

\noindent which defines the atom number enhancement $\mathcal{E}$ due to the repumping beam. We furthermore note that in steady state $N_3^{ss} = \frac{\Gamma_{23}}{\Gamma_{34}+\Gamma_{36}} N_2^{ss}$. Using the decay rates given in Ref. \citep{xu2003} this means that in steady-state conditions there are about twice as many atoms in the invisible transient state $|3\rangle$ in the system than in the fluorescescing state $|2\rangle$.

We can express the pumping rate on the repumping transition as $\beta = \frac{\sigma}{\hbar\omega} I$, with $\sigma$ as the absorption cross section in the lambda level scheme
\begin{eqnarray}\label{eq:RepPumpingRate}
\sigma = \frac{\Gamma_{54}}{\Gamma_5} \frac{6\pi c^2}{\omega^2} \frac{\Gamma_5^2}{4\Delta^2+\Gamma_5^2} \equiv \sigma_0 \frac{\Gamma_5^2}{4\Delta^2+\Gamma_5^2}~.
\end{eqnarray}

\noindent Here, $c$ is the speed of light, $\omega$ the angular frequency of the repumping transition, $I$ the intensity, and $\Delta$ the detuning of the repumping beam, The total decay rate from the doubly-excited state is given by $\Gamma_5 = \Gamma_{54} + \Gamma_{56}$. The prefactor $\frac{\Gamma_{54}}{\Gamma_5}$ stems from the additional decay channel $\Gamma_{56}$. Under the condition $\Gamma_{4x} \ll \Gamma_{5x} \ll \Gamma_{56}$, we can write the enhancement as
\begin{eqnarray}
\mathcal{E} = 1 + \mathcal{E}_0 \, \frac{w^2}{4\Delta^2 + w^2}~,
\end{eqnarray}

\noindent which is a Lorentzian with the amplitude
\begin{eqnarray}\label{eq:ampl}
\mathcal{E}_0 = \frac{\Gamma_{56}}{\Gamma_{5x}} \frac{s}{1 + s}~,
\end{eqnarray}

\noindent with $s = I/I_s$ denoting the saturation parameter, and the width
\begin{eqnarray}\label{eq:width}
w = \Gamma_5 \sqrt{1+s}~.
\end{eqnarray}

\noindent The effective saturation intensity in the $\Lambda$ scheme, considering both the relevant transitions $5\mathrm{p}^2 \, ^3\mathrm{P}_2 \rightarrow 5\mathrm{s}5\mathrm{p} \, ^3\mathrm{P}_{1,2}$ and losses, is defined as
\begin{eqnarray}\label{eq:Isat}
I_s = \frac{\Gamma_{4x}}{\Gamma_{5x}} \frac{\omega}{\sigma_0} \, \hbar \, \Gamma_5~.
\end{eqnarray}

\noindent Eqs. (\ref{eq:ampl}) and (\ref{eq:width}) are used to fit the experimental data in Fig. \ref{fig:fig2} (b).

If we allow for a weak decay $\Gamma_{2x}$ from the fluorescing state instead of the weak decay from the doubly-excited state $\Gamma_{5x}$, and furthermore using the approximations $\Gamma_{4x} \ll \Gamma_{56}$ and $\Gamma_{2x} \ll \Gamma_{23}$, then all the equations above are still valid, except that $\Gamma_{5x}$ is replaced by $\Gamma_{5x} = \frac{\Gamma_{34} + \Gamma_{36}}{\Gamma_{34}} \frac{\Gamma_{56}}{\Gamma_{23}} \Gamma_{2x}$ in Eq. (\ref{eq:ampl}) and Eq. (\ref{eq:Isat}). This means that from our experiment the decay mechanism can not be unambiguously assigned to either $\Gamma_{5x}$ or $\Gamma_{2x}$.

\bibliography{repump_references}

\end{document}